\documentclass[12pt]{article}

\usepackage{amsmath}
\usepackage{latexsym}
\usepackage{bbm}

\def\a{\alpha}

\def\b{\beta}
\def\g{\gamma}

\def\d{\delta}

\def\e{\varepsilon}

\def\m{\mu}

\def\r{\rho}

\def\s{\sigma}
\def\t{\tau}

\def\G{\Gamma}

\def\pa{\partial}

\def\half{\frac{1}{2}}

\def\IZ{\mathbbm Z}

\def\half{\frac{1}{2}}

\def\and{{\rm and}}

\def\ie{{\it i.e.,} }

\begin{document}

%\vspace*{-.6in} \thispagestyle{empty}
%\begin{flushright}
%CALT-68-??
%\end{flushright}
\baselineskip = 16pt

\vspace{1.5in} {\Large
\begin{center}
THE EARLY HISTORY OF\\ STRING THEORY AND SUPERSYMMETRY
\end{center}}

%\vspace{.5in}

\begin{center}
John H. Schwarz
\\
\emph{California Institute of Technology\\ Pasadena, CA  91125, USA}
\end{center}

%\vspace{1in}

\begin{center}
\textbf{Abstract}
\end{center}
\begin{quotation}
\noindent This lecture presents a brief overview of the early history
of string theory and supersymmetry. It describes how the S-matrix theory
program for understanding the strong nuclear force
evolved into superstring theory, which is a promising framework for constructing a
unified quantum theory of all forces including gravity. The period covered begins
with S-matrix theory in the mid 1960s and ends with the widespread acceptance of
superstring theory in the mid 1980s. Further details and additional
references can be found in Schwarz (2007).

\end{quotation}

%\newpage

\section{S-Matrix Theory}

In UC Berkeley, where I was a graduate student in the mid 1960s,
Geoffrey Chew (my thesis advisor), Stanley Mandelstam, and others
focussed their efforts on constructing a theory of the strong
nuclear force, \ie a theory of hadrons.
Chew's approach to understanding the strong nuclear force
was based on {\em S-matrix theory}. He argued that
quantum field theory, which was so successful in describing
QED, was inappropriate for describing a strongly interacting theory,
where a weak-coupling perturbation expansion would not be useful. One
reason for holding this view was that none of the hadrons seemed more
fundamental than any of the others. Therefore a field theory that
singled out some subset of the hadrons did not seem sensible. Also,
it seemed impossible to formulate a quantum field theory with
a fundamental field for every hadron. Chew spoke of {\em nuclear
democracy} and the {\em bootstrap principle} to describe this situation.
Chew advocated focussing attention on physical quantities, especially the
S Matrix, which describes on-mass-shell scattering amplitudes. The goal
was to develop a theory that would determine the hadron spectrum
and hadronic S matrix.

The quark concept also arose during this period,
but the prevailing opinion in the mid 1960s was that quarks
are mathematical constructs, rather than physical entities,
whose main use is as a mathematical technique for understanding
symmetries and quantum numbers. The SLAC deep inelastic scattering
experiments in the late 1960s made it clear that quarks and gluons
are physical (confined) particles. It was then natural to try to
base a quantum field theory on them, and QCD was developed a few
years later with the discovery of asymptotic freedom. Thus, with the
wisdom of hindsight, it is clear that Chew {\it et al.} were wrong to
reject quantum field theory. Nonetheless, their insights were
very influential, perhaps even crucial, for the discovery of
string theory, which can be regarded as the ultimate realization of
the S-Matrix program.

Some of the ingredients that went into the
S-matrix theory program, such as unitarity and maximal analyticity
of the S matrix, were properties (deduced
from quantum field theory) that encode the requirements of
causality and nonnegative probabilities.
Another important ingredient was analyticity in angular
momentum. The idea is that partial
wave amplitudes $a_l (s)$, which are defined in the first instance
for angular momenta $l=0,1,\ldots$, can be extended to an
analytic function of $l$, $a(l,s)$. The uniqueness of this extension
results from imposing suitable asymptotic behavior in $l$.
The Mandelstam invariant $s$
is the square of the center-of-mass energy of the scattering reaction.
The analytic function $a(l,s)$ can have isolated poles called {\em Regge
poles}. (Branch points are also possible, but they are usually ignored.)
The position of a Regge pole is
given by a {\em Regge trajectory} $l = \a(s)$. A value of $s$ for which
$l = \a(s)$ takes a physical value corresponds to a physical hadron of spin $l$.

Theoretical work in this period was strongly influenced by
experimental results. Many new hadrons were discovered in
experiments at the Bevatron in Berkeley, the AGS in Brookhaven, and
the PS at CERN. Plotting masses squared versus angular momentum (for
fixed values of other quantum numbers), it was noticed that the
Regge trajectories are approximately linear with a common slope
$$
\a(s) = \a(0) + \a' s \qquad \qquad \a' \sim 1.0\, ({\rm
GeV})^{-2}\, .
$$
Using the crossing-symmetry properties of analytically continued
scattering amplitudes, one argued that exchange of Regge poles (in
the $t$ channel) controlled the high-energy, fixed momentum
transfer, asymptotic behavior of physical amplitudes:
$$
A(s, t) \sim \b(t) (s/s_0)^{\a(t)} \qquad s \to \infty, \, t <0.
$$
In this way one deduced from data that the intercept of the $\r$
trajectory, for example, was $\a_\r(0) \sim .5$. This is consistent
with the measured mass $m_\r =.76 \, {\rm GeV}$ and the Regge slope
$\a' \sim 1.0\, ({\rm GeV})^{-2}$.

The approximation of linear
Regge trajectories describes long-lived resonances, whose widths
are negligible compared to their masses. This approximation is called the
{\em narrow resonance approximation}. In
this approximation branch cuts in scattering amplitudes, whose
branch points correspond to multiparticle thresholds, are
approximated by a sequence of resonance poles. This is what one
would expect in the tree approximation to a quantum field theory
in which all the resonances appear as fundamental fields.
However, there was also another discovery, called {\em duality},
which clashed with the usual notions of quantum field theory.
In this context duality means
that a scattering amplitude can be expanded in an infinite series
of $s$-channel poles, and this gives the same result as its
expansion in an infinite series of $t$-channel poles. To include both
sets of poles, as usual Feynman diagram techniques might suggest,
would amount to double counting.

\section{The Discovery of String Theory}

Veneziano (1968)
discovered a simple analytic formula that exhibits
duality with linear Regge trajectories.
It is given by a sum of ratios of Euler gamma functions:
$$
T = A(s,t) + A(s,u) + A(t,u), \quad {\rm where} \quad
A(s,t) = g^2\frac{\G(-\a(s)) \G(- \a(t))}{\G(-\a(s) -\a(t)) },
$$
$g$ is a coupling constant, and $\a$ is a linear Regge trajectory
$$
\a(s) = \a(0) + \a' s .
$$
The Veneziano formula gives an explicit realization of duality and
Regge behavior in the narrow resonance approximation. The
function $A(s,t)$ can be expanded as an infinite series of $s$-channel poles
or of $t$-channel poles. The motivation for writing down this
formula was largely phenomenological, but it turned out that formulas
of this type describe scattering amplitudes in the tree approximation
to a consistent quantum theory!

A generalization to incorporate adjoint
$SU(N)$ quantum numbers was formulated by Paton and Chan
(1969). Chan--Paton symmetry was initially envisaged to be a
global (flavor) symmetry, but it was shown later to be a local
gauge symmetry.

Very soon after the appearance of the Veneziano amplitude, Virasoro (1969)
proposed an alternative formula
$$
T = g^2\frac{\G(-\half \a(s)) \G(- \half\a(t))
\G(- \half\a(u))}{\G(-\half\a(t) -\half\a(u)) \G(-\half\a(s)
-\half\a(u))\G(-\half\a(s) -\half\a(t)) },
$$
which has similar virtues. Since this formula has total $stu$
symmetry, it describes particles that are singlets of
the Chan--Paton symmetry group.

Over the course of the next year or so, {\em dual
models}, as the subject was then called, underwent a sudden surge of
popularity, marked by several remarkable discoveries. One was the
discovery (by several different groups) of an $N$-particle generalization
of the Veneziano formula
$$
A_N(k_1,k_2,\dots,k_N) = g_{\rm open}^{N-2} \int d\mu_N(y)  \prod_{i<j} (y_i -
y_j)^{\a' k_i \cdot k_j},
$$
where $y_1, y_2, \ldots, y_N$ are real coordinates. I will omit the description
of the measure $d\mu_N(y)$, which can be found in Schwarz (2007).
This formula has cyclic symmetry in the $N$ external lines.
Soon thereafter Shapiro (1970) formulated an $N$-particle generalization of
the Virasoro formula:
$$
A_N(k_1,k_2,\dots,k_N) = g_{\rm closed}^{N-2} \int
d\mu_N(z)\prod_{i<j} |z_i-z_j|^{\a'k_i\cdot k_j } ,
$$
where $z_1, z_2, \ldots, z_N$  are complex coordinates. This amplitude has
total symmetry in the $N$ external lines.

Both of these formulas for multiparticle amplitudes
were shown to have poles whose residues factorize in a consistent manner
on an infinite spectrum of single-particle states. This spectrum is described
by a Fock space associated to an infinite number of harmonic oscillators
$$ \{ a^\m_m \} \qquad \m = 0,1,\dots,d-1 \qquad m= 1, 2, \ldots $$
where $d$ is the dimension of Minkowski spacetime, which was initially assumed
to be four.
There is one set of such oscillators in the Veneziano case and two sets
in the Shapiro--Virasoro case. These spectra were interpreted as describing
the normal modes of a relativistic string:
an open string (with ends) in the first case and a closed string (loop)
in the second case. Amazingly, the formulas were discovered
before this interpretation was proposed. In the above formulas, the
$y$ coordinates parametrize points on the boundary of a string world sheet,
where particles that are open-string states are emitted or absorbed,
whereas the $z$ coordinates parametrize points on the interior of a
string world sheet, where particles that are closed-string states are
emitted or absorbed. (It is also possible to construct amplitudes in which
both types of particles participate.)

Having found the factorization, it became possible to compute
radiative corrections (loop amplitudes). Gross, Neveu, Scherk, and
Schwarz (1970) discovered unanticipated
singularities in a particular one-loop diagram for which
the world sheet is a cylinder with two external particles attached
to each of the two boundaries. The computations showed that
this diagram gives branch points that violate unitarity. This was a
very disturbing conclusion, since it seemed to imply that the
classical theory does not have a consistent quantum extension.
However, soon thereafter it was pointed out by Lovelace (1971)
that these branch points become poles provided that
$$
\a(0)=1 \quad \and \quad d=26.
$$
Prior to this discovery, everyone assumed that the spacetime dimension
should be $d=4$. We had no physical reason to consider extra dimensions.
It was the mathematics that forced us in that direction. Later,
these poles were interpreted as closed-string states in a one-loop
open-string amplitude. Nowadays this is referred to as
{\em open-string/closed-string duality}. This is closely related
to {\em gauge/gravity duality}, which was discovered 27 years later.

The analysis also required there to be an infinite number of
decoupling conditions, which turned out to coincide with the
constraints proposed by Virasoro (1970) and further elaborated
upon by Fubini and Veneziano (1971).
Since the string has an infinite spectrum of higher-spin states, there
are corresponding gauge invariances that eliminate unphysical degrees
of freedom. The operators that describe the constraints that arise
for a particular covariant gauge choice satisfy the Virasoro algebra
$$ [L_m, L_n] = (m-n) L_{m+n} + \frac{c}{12} (m^3 - m)\d_{m,-n},$$
where $m,n$ are arbitrary integers. These operators
can also be interpreted as generators of conformal symmetry for the
two-dimensional string world sheet. The central charge (or conformal anomaly)
$c$ is equal to the spacetime dimension $d$. This anomaly cancels
for $d=26$ when the contribution of Faddeev--Popov ghosts is included.

\section{The RNS Model and the Discovery of Supersymmetry}

In a very inspired and important development,
Ramond (1971) constructed a stringy analog of the Dirac equation,
which describes a fermionic string. Just as
the string momentum $p^\m$ is the zero mode of a
density $P^\m (\s)$, where the coordinate $\s$ parametrizes
the string, he proposed that the Dirac matrices $\g^\m$ should be the
zero modes of densities $\G^\m (\s)$.  Then he considered the
Fourier modes of the dot product:
$$ F_n = \int_0^{2\pi} e^{-in\s} \G (\s)\cdot P (\s)d\s \qquad n\in \IZ.$$
In particular,
$$ F_0 = \g \cdot p + {\rm additional \, terms}.$$
He proposed that physical states of a fermionic string should satisfy
the following analog of the Dirac equation
$$ (F_0 + M) |\psi\rangle =0.$$
He also observed that in the case of the fermionic string
the Virasoro algebra generalizes to a super-Virasoro algebra
$$ \{ F_m, F_n \} = 2 L_{m+n} + \frac{c}{3} m^2 \d_{m,-n}$$
$$ [ L_m, F_n ] = (\frac{m}{2} -n) F_{m+n}$$
$$ [L_m, L_n] = (m-n) L_{m+n} + \frac{c}{12} m^3 \d_{m,-n}.$$
Ramond's paper does not include the central extension, which turns out
to be $c=3d/2$, where $d$ is the spacetime dimension. A little later, it
was realized that consistency requires $d=10$ and $M=0$.
These conditions are the analogs of $d=26$ and $\a(0)=1$ for the
bosonic Veneziano string theory.

A couple of months later Neveu and Schwarz (1971a) constructed
a new interacting bosonic string theory,
which was called the {\em dual pion model}.
It has a similar structure to the fermionic string,
but the periodic density $\G^\m(\s)$ is
replaced by an antiperiodic one $H^\m (\s + 2\pi) = - H^\m(\s)$.
Then the Fourier modes, which differ from an integer by 1/2,
$$ G_r = \int_0^{2\pi} e^{-ir\s} H \cdot P d\s \qquad r \in \IZ + 1/2$$
satisfy a similar super-Virasoro algebra. Neveu and Schwarz (1971a) refers to
this algebra as a {\em supergauge algebra}, a terminology that was sensible in the
context at hand. The Neveu--Schwarz bosons and Ramond fermions were combined
in a unified interacting theory of bosons and fermions by Neveu and Schwarz (1971b)
and by Thorn (1971). This theory (the RNS model)
was an early version of superstring theory. As will be
explained shortly, a few crucial issues were not yet understood.

After a few more months, Gervais and Sakita (1971)
showed that the the RNS model is described
by the string world-sheet action
$$ S = T\int d\s d\t \left( \pa_\a X^\m \pa^\a X_\m -i \bar\psi^\m \g^\a
\pa_\a \psi_\m\right),$$
where the coefficient $T$ is the string tension. They also explained
that it has {\em two-dimensional supersymmetry}, though that terminology was not used yet,
by showing that it is invariant under the transformations
$$\d X^\m = \bar\e \psi^\m, \quad \quad \d\psi^\m = -i \g^\a \e \pa_\a X^\m ,$$
where $\e$ is an infinitesimal constant spinor. To the best of my knowledge,
this is the first supersymmetric theory identified in the literature!
There are two possibilities for the world-sheet fermi
fields $\psi^\m$. When it is antiperiodic $\psi^\m = H^\m$, which gives the boson spectrum
(Neveu--Schwarz sector), and when it is periodic $\psi^\m = \G^\m$, which gives the fermion
spectrum (Ramond sector).

Five years later, Brink, Di Vecchia, and Howe (1976) and Deser and Zumino (1976)
constructed a more fundamental world-sheet action with
local supersymmetry. This formulation of the world-sheet theory has
the additional virtue of also accounting for the super-Virasoro
constraints. From this point of view,
the significance of the super-Virasoro algebra is that the
world-sheet theory, when properly gauge fixed and quantized, has
{\em superconformal symmetry}. Again, the anomaly cancels for $d=10$
when the Faddeev--Popov ghosts are included.

At about the same time as Ramond's paper, the four-dimensional super-Poincar\'e
algebra was introduced in a paper by Golfand and Likhtman (1971), who
proposed constructing 4d field theories with this symmetry. This
paper went unnoticed in the West for several more years. In fact,
the celebrated paper of Wess and Zumino (1974), which formulated a
class of 4d supersymmetric theories, was motivated by the search for 4d analogs of the
2d Gervais--Sakita world-sheet action. The Wess-Zumino paper launched the study
of supersymmetric field theories, which proceeded in parallel
with the development of supersymmetric string theory. Wess and Zumino (1974) used
the expression {\em supergauge}, following the terminology of Neveu and Schwarz (1971),
but in their subsequent papers they switched to {\em supersymmetry},
which was more appropriate for what they were doing.

\section{The Temporary Demise of String Theory}

String theory is formulated as an on-shell S-matrix theory in
keeping with its origins discussed earlier. However, the SLAC deep
inelastic scattering experiments in the late 1960s made it clear
that the hadronic component of the electromagnetic current is a
physical off-shell quantity, and that its asymptotic properties
imply that hadrons have hard pointlike constituents. Moreover,
all indications (at that time) were that strings
are too soft to describe hadrons with their pointlike constituents.

By 1973--74 there were many good reasons to stop working on string
theory: a successful and convincing theory of hadrons (QCD) was
discovered, and string theory had severe problems as a theory of hadrons.
These included an unrealistic spacetime dimension ($d=10$ or $d=26$),
an unrealistic spectrum (including a tachyon and massless particles),
and the absence of pointlike constituents. A few years of attempts
to do better had been unsuccessful.
Moreover, convincing theoretical and experimental evidence for the
Standard Model was rapidly falling into place. That was where the
action was. Even for those seeking to pursue speculative theoretical ideas
there were options other than string theory
that most people found more appealing, such as
grand unification and supersymmetric field theory. Understandably,
string theory fell out of favor. What had been a booming enterprise
involving several hundred theorists rapidly came to a grinding halt.
Only a few diehards continued to pursue it.

\section{Gravity and Unification}

Among the problems of the known string theories, as a theory of
hadrons, was the fact that the spectrum of open strings contains
massless spin 1 particles, and the spectrum of closed strings
contains a massless spin 2 particle (as well as other massless
particles), but there are no massless hadrons. In 1974,
Jo\"el Scherk and I decided to take string theory seriously as it stood, rather
than forcing it to conform to our preconceptions. This meant abandoning the original
program of describing hadron physics and interpreting the massless spin 2 state
in the closed-string spectrum as a graviton. Also, the massless spin 1 states
in the open-string spectrum could be interpreted as particles
associated to Yang--Mills gauge fields. Specifically,
Scherk and Schwarz (1974) proposed trying to interpret string theory as a
unified quantum theory of all forces including gravity. Neveu and Scherk (1972) had
shown that string theory
incorporates the correct gauge invariances to ensure agreement at low energies
(compared to the scale given by the string tension) with Yang--Mills
theory. Yoneya (1973,1974) and Scherk and Schwarz (1974) showed that it also
contains gauge invariances that
ensure agreement at low energies with general relativity.

To account for Newton's constant, the most natural choice for the fundamental
string length scale was $l_s \sim 10^{-33}$ cm  (the Planck length) instead of
$l_s \sim 10^{-13} $ cm (the typical size of a hadron). Thus the strings suddenly
shrank by 20 orders of magnitude, but the mathematics was
essentially unchanged. The string tension is proportional to
$l_s^{-2}$, so it increased by 40 orders of magnitude.

The proposed new interpretation had several advantages:

\noindent $\bullet$ Gravity and Yang--Mills forces are required by string theory.

\noindent $\bullet$ String theory has no UV divergences.

\noindent $\bullet$ Extra spatial dimensions could be a good thing.

\noindent Let me say a few words about the last point. In a nongravitational theory, the
spacetime geometry is a rigid background on which the dynamics takes place. In that
setup, the fact that we observe four-dimensional Minkowski spacetime is a
compelling argument to formulate the theory in that background geometry.
As you know very well, this is part of the story of the Standard Model. However, in
a gravitational theory that abides by the general principles laid out by Einstein,
the spacetime geometry is determined by the dynamical equations. In such a setup extra
dimensions can make sense provided that the equations of the theory have a solution
for which the geometry is the product of four-dimensional Minkowski spacetime and
a compact manifold that is sufficiently small to have eluded detection. It
turns out that there are many such solutions. Moreover, the details of the
compact manifold play a crucial role in determining the symmetries and particle
content of the effective low-energy theory in four dimensions,
even when the compact dimensions are much too small to observe directly.

\section{Supersymmetry, Supergravity, and Superstrings}

In the second half of the 1970s the study of supersymmetric field
theories become a major endeavor.
A few important supersymmetric theories that were formulated in that era included

\noindent $\bullet$ ${\cal N} = 1$, $d=4$ supergravity, discovered by
Freedman, Van Nieuwenhuizen, and Ferrara (1976) and Deser and Zumino (1976).

\noindent $\bullet$ ${\cal N} = 1$, $d=10$ and ${\cal N} = 4$, $d=4$
supersymmetric Yang--Mills theory discovered by Brink, Scherk, and Schwarz (1977)
and Gliozzi, Scherk, and Olive (1977).

\noindent $\bullet$ ${\cal N} = 1$, $d=11$ supergravity discovered by
Cremmer, Julia and Scherk (1978).

Gliozzi, Scherk, and Olive (1976, 1977) proposed a truncation of the RNS string
theory spectrum -- {\em the GSO Projection} -- that
removes half of the fermion states and the ``odd G-parity'' bosons.
In particular, the latter projection eliminates the tachyon.
They showed that after the projection the number of physical bosonic
degrees of freedom is equal to the number of physical
fermionic degrees of freedom at every mass level.
This was compelling evidence for {\em ten-dimensional spacetime
supersymmetry} of the GSO-projected theory. Prior to this, we knew
about the supersymmetry of the two-dimensional string world-sheet theory,
but we had not considered the possibility of spacetime supersymmetry. In fact,
the GSO projection is not just an option; it is required for consistency.

In 1979 Michael Green and I began a collaboration, which had the initial goal of
understanding and proving the ten-dimensional
spacetime supersymmetry of the GSO-projected version of the RNS theory.
The highlights of our work included Green and Schwarz (1981, 1984a), which
developed a new formalism in which the spacetime supersymmetry of the GSO-projected
RNS string is manifest, and Green and Schwarz (1982), which classified the
consistent ten-dimensional superstring theories and giving them the names
Type I, Type IIA, and Type IIB.
We were excited about these (and other) developments, but they did not
arouse much interest in the theory community. String theory was still in the
doldrums.

In the early 1980s
there was growing interest in supersymmetry and extra dimensions.
In particular, a small community became intrigued by Kaluza--Klein reduction
of 11-dimensional supergravity. Only the string ingredient was missing from
their considerations. That changed following our next discovery.

\section{Anomalies}

If a unified theory is to make contact with the Standard Model, and
have a chance of being realistic, parity violation is an essential ingredient.
However, parity-violating classical theories generically have
{\em gauge anomalies}, which means that they cannot be used to
define quantum theories. The gauge symmetry is broken by
one-loop quantum corrections, rendering the would-be quantum theory
inconsistent. In the case of the Standard Model, if one were to
change the theory by removing
all of the leptons or all of the quarks, the theory would become inconsistent.
When both the quarks and the leptons are
included all gauge anomalies beautifully cancel, and so the Standard Model is a
well-defined quantum theory. These considerations raise the question whether
the potential gauge anomalies in chiral superstring
theories also cancel, so that they give consistent quantum theories.

We knew that Type I superstring theory
is a well-defined ten-dimensional theory at tree level for any $SO(n)$
or $Sp(n)$ gauge group, and that for every such group
it is chiral (\ie parity violating). However, evaluation of a one-loop hexagon
diagram in ten-dimensional super Yang--Mills theory, which describes
the massless open-string states, exhibits explicit nonconservation of
gauge currents, signalling a gauge anomaly. The only
hope for consistency is that inclusion of the closed-string (gravitational)
sector cancels this gauge anomaly without introducing new ones.

Type IIB superstring theory, which only has a closed-string gravitational
sector, is also chiral and therefore potentially anomalous. It was not
known how to analyze such anomalies until
Alvarez-Gaum\'e and Witten (1984) derived general formulas
for gauge, gravitational, and mixed anomalies in an arbitrary spacetime
dimension. Using their results, they discovered that the
gravitational anomalies, which would imply nonconservation of the stress tensor,
cancel in Type IIB superstring theory. In their calculation this cancellation
appears quite miraculous,
though the UV finiteness of the Type IIB loop amplitudes implies that it had to work.
Thus, Type IIB is a consistent chiral superstring theory.
On the other hand, it did not look promising
for describing the real world, since it does not contain any Yang--Mills gauge
fields. (Many years later, nonperturbative Type IIB solutions that do contain
Yang--Mills fields were discovered.)
At that time, the last hope for constructing a realistic model
seemed to reside with the Type I superstring
theories, which are chiral and do contain Yang--Mills fields.

After a couple years of failed attempts, Green and I finally managed
to compute the one-loop hexagon diagrams in Type I superstring theory.
We found that both the cylinder and the M\"obius-strip
world-sheet diagrams contribute to the gauge anomaly and realized
that there might be a gauge group for which the
two contributions cancel. Green and Schwarz (1985) showed that
$SO(32)$ is the unique choice for which the cancellation
occurs. Since this computation
only demonstrated the cancellation of the pure gauge part of the anomaly,
we decided to explore the low-energy effective field theory to see
whether the gravitational and mixed anomalies also cancel.
Using the results of Alvarez-Gaum\'e and Witten (1984), Green and Schwarz (1984b)
verified that all gauge, gravitational, and mixed anomalies do in fact cancel
for the gauge group $SO(32)$.

The effective field theory analysis showed that $ E_8 \times E_8$ is
a second (and the only other) gauge group for which the anomalies could cancel
for a theory with ${\cal N} =1 $ supersymmetry in ten dimensions. In both
cases, it is crucial for the result that the coupling to
supergravity is included. The $SO(32)$ case could be accommodated by
Type I superstring theory, but we didn't know a superstring theory
with gauge group $E_8 \times E_8.$ We were aware of the article by
Goddard and Olive (1983) that pointed out (among other things) that there
are exactly two even self-dual Euclidean lattices in 16 dimensions, and
these are associated with precisely these two gauge groups.
However, we did not figure out how to exploit this fact before the problem
was solved by Gross, Harvey, Martinec, and Rohm (1985).

\section{Epilogue}

Following these discoveries there was a sudden surge of interest in
superstring theory. After more than a decade, string theory had emerged
from the doldrums.
In my view, some of the new converts made a phase transition from being
too pessimistic about string theory to being
too optimistic about the near-term prospects for finding a realistic
model. However, after a few years, almost all practitioners
had a much more sober assessment of the challenges that remain.
Superstring theory (including M-theory, which is part of the same
theoretical framework) has remained a very active subject ever
since 1984. Even though the construction of a complete and realistic model
of elementary particles still appears to be a distant dream, the
study of string theory has been enormously productive. For example,
insights derived from these studies have had a profound impact on
fundamental mathematics and are beginning to inspire new approaches
to understanding topics in other areas of physics.

For many years string theory was
considered to be a radical alternative to quantum field theory.
However, in recent times -- long after the period
covered by this lecture -- dualities relating string theory
and quantum field theory were discovered. In view of these dualities,
my current opinion is that string theory is best regarded as the
logical completion of quantum field theory, and therefore it is not
radical at all. There is still much that remains to be understood,
but I am convinced that we are on the right track and making very
good progress.

This work was supported in part by the U.S. Dept. of
Energy under Grant No. DE-FG03-92-ER40701.

\newpage

\def \AP {{\it Annals of Physics}, }
\def \NC {{\it Nuovo Cimento A}, }
\def \NCL {{\it Nuovo Cimento Letters}, }
\def \NP {{\it Nuclear Physics B}, }
\def \PL {{\it Physics Letters B}, }
\def \PR {{\it Physical Review D}, }
\def \PRL {{\it Physical Review Letters}, }
\def \PTP {{\it Progress in Theoretical Physics}, }

\baselineskip = 24pt

\centerline{References}

\noindent Alvarez-Gaum\'e, L. \& Witten, E. (1984).
Gravitational anomalies.
\NP \\
\hspace*{.5in}{\it 234}, 269-330.

\noindent Brink, L., Di~Vecchia, P., \& Howe, P. (1976).
A locally supersymmetric and\\
\hspace*{.5in} reparametrization invariant action for the spinning string.
\PL\\
\hspace*{.5in}{\it 65}, 471-474.

\noindent Brink, L., Schwarz, J. H., \& Scherk, J. (1977).
Supersymmetric Yang-Mills theories.\\
\hspace*{.5in}\NP {\it 121}, 77-92.

\noindent Cremmer, E., Julia, B., \& Scherk, J. (1978).
Supergravity theory in eleven-dimensions.\\
\hspace*{.5in}\PL {\it 76}, 409-412.

\noindent Deser, S., \& Zumino, B. (1976a).
Consistent supergravity.
\PL {\it 62}, 335-337.

\noindent Deser, S., \& Zumino, B. (1976b).
A complete action for the spinning string. \\
\hspace*{.5in}\PL {\it 65}, 369-373.

\noindent Freedman, D. Z., Van Nieuwenhuizen, P., \& Ferrara, S. (1976).
Progress toward a theory\\
\hspace*{.5in}of supergravity.
\PR {\it 13}, 3214-3218.

\noindent Fubini, S., \& Veneziano, G.~(1971).
Algebraic treatment of subsidiary conditions in dual\\
\hspace*{.5in}resonance models.
\AP {\it 63}, 12-27.

\noindent Gervais, J. L., \& Sakita, B. (1971).
Field theory interpretation of supergauges in dual\\
\hspace*{.5in}models.
\NP {\it 34}, 632-639.

\noindent Gliozzi, F., Scherk, J., \& Olive, D. (1976).
Supergravity and the spinor dual model.\\
\hspace*{.5in}\PL {\it 65}, 282-286.

\noindent Gliozzi, F., Scherk, J., \& Olive, D. (1977).
Supersymmetry, supergravity theories and the\\
\hspace*{.5in}dual spinor model.
\NP {\it 122}, 253-290.

\noindent Goddard, P., \& Olive, D. (1983).
Algebras, lattices and strings. DAMTP-83/22. Reprinted\\
\hspace*{.5in}in
Goddard, P. (Ed.), \& Olive, D. (Ed.): {\it Kac-Moody and Virasoro Algebras}\\
\hspace*{.5in}(pp. 210-255). World Scientific 1988.

\noindent Golfand, Yu.~A., \& Likhtman, E.~P.~(1971).
Extension of the algebra of Poincar\'e group\\
\hspace*{.5in}generators and violation of P invariance.
{\it JETP Lett.}, {\it 13}, 323-326.

\noindent Green, M.~B., \& Schwarz, J.~H.~(1981).
Supersymmetrical dual string theory.
{\it Nuclear \\ \hspace*{.5in}Physics B}, {\it 181}, 502-530.

\noindent Green, M.~B., \& Schwarz, J.~H.~(1982).
Supersymmetrical string theories.
{\it Physics\\ \hspace*{.5in} Letters B}, {\it 109}, 444-448.

\noindent Green, M.~B., \& Schwarz, J.~H.~(1984a).
Covariant description of superstrings.
{\it Physics\\ \hspace*{.5in} Letters B}, {\it 136}, 367-370.

\noindent Green, M.~B., \& Schwarz, J.~H.~(1984b).
Anomaly cancellation in supersymmetric $D=10$\\
\hspace*{.5in}gauge theory and superstring theory.
\PL {\it 149}, 117-122.

\noindent Green, M.~B., \& Schwarz, J.~H.~(1985).
The hexagon gauge anomaly in Type I superstring\\
\hspace*{.5in}theory.
\NP {\it 255}, 93-114.

\noindent Gross, D.~J., Neveu, A., Scherk, J., \& Schwarz, J.~H.~(1970).
Renormalization and unitarity\\
\hspace*{.5in}in the dual resonance model.
\PR {\it 2}, 697-710.

\noindent Gross, D.~J., Harvey, J.~A., Martinec, E.~J., \& Rohm, R.~(1985).
The heterotic string.\\
\hspace*{.5in}\PRL {\it 54}, 502-505.

\noindent Lovelace, C. (1971).
Pomeron form factors and dual Regge cuts.
\PL {\it 34}, 500-506.

\noindent Neveu, A., \& Schwarz, J.~H.~(1971a).
Factorizable dual model of pions.
\NP\\
\hspace*{.5in}{\it 31}, 86-112.

\noindent Neveu, A., \& Schwarz, J.~H.~(1971b).
Quark model of dual pions
\PR {\it 4},\\
\hspace*{.5in}1109-1111.

\noindent Neveu, A., Schwarz, J.~H., \& Thorn, C.~B.~(1971).
Reformulation of the dual pion\\
\hspace*{.5in}model.
\PL {\it 35}, 529-533.

\noindent Neveu, A., \&  Scherk, J. (1972).
Connection between Yang--Mills fields and dual models.\\
\hspace*{.5in}\NP {\it 36}, 155-161.

\noindent Paton, J.~E.,~\& Chan, H.~(1969).
Generalized Veneziano model with isospin.
{\it Nuclear \\ \hspace*{.5in}Physics B}, {\it 10}, 516-520.

\noindent Ramond, P. (1971).
Dual theory for free fermions.
\PR {\it 3}, 2415-2418.

\noindent Scherk, J., \& Schwarz, J.~H.~(1974).
Dual models for non-hadrons.
\NP\\
\hspace*{.5in}{\it 81}, 118-144.

\noindent Schwarz, J.~H.~(2007).
The early years of string theory: a personal perspective.\\
\hspace*{.5in}arXiv:0708.1917 [hep-th].

\noindent Shapiro, J.~A.~(1970).
Electrostatic analog for the Virasoro model.
\PL {\it 33},\\
\hspace*{.5in}361-362.

\noindent Thorn, C.~B.~(1971).
Embryonic dual model for pions and fermions.
\PR\\
\hspace*{.5in}{\it 4}, 1112-1116.

\noindent Veneziano, G.~(1968).
Construction of a crossing-symmetric Regge-behaved
amplitude for\\
\hspace*{.5in}linearly rising Regge trajectories.
\NC {\it 57}, 190-197.

\noindent Virasoro, M.~(1969).
Alternative constructions of crossing-symmetric
amplitudes with\\
\hspace*{.5in}Regge behavior.
{\it Physical Review}, {\it 177}, 2309-2311.

\noindent Virasoro, M.~(1970).
Subsidiary conditions and ghosts in dual resonance models.\\
\hspace*{.5in}\PR {\it 1}, 2933-2936.

\noindent Wess, J., \& Zumino, B.~(1974).
Supergauge transformations in four dimensions.
{\it Nuclear \\ \hspace*{.5in}Physics B}, {\it 70}, 39-50.

\noindent Yoneya, T.~(1973).
Quantum gravity and the zero slope limit of the generalized Virasoro\\
\hspace*{.5in}model.
\NCL {\it 8}, 951-955.

\noindent Yoneya, T.~(1974).
Connection of dual models to electrodynamics and gravidynamics.\\
\hspace*{.5in}\PTP {\it 51}, 1907-1920.

\end{document}